# Técnicas Monte Carlo para la enseñanza de la estadística

*Bueno Pérez, F.M. Alexander y Manzano Diosdado, Daniel*

Instituto Carlos I de Física Teórica y Computacional

**Resumen**

La probabilidad y la estadística son una de las ramas de las matemáticas con más aplicaciones, y cuyos conceptos son más difíciles de asimilar. Mientras en otras ramas, como las ciencias naturales, la experimentación suele ser una herramienta para desarrollar la intuición de los estudiantes. En el estudio de la probabilidad este desarrollo es más difícil. En este artículo mostramos como las técnicas Monte Carlo y la realización de experimentos por ordenador con números pseudoaleatorios, puede ser muy útil en la asimilación de estas materias. Estas técnicas son muy usadas en investigación científica, pero su enseñanza suele estar relegada a cursos específicos de educación superior. Mediante la simulación por ordenador se puede también facilitar mucho este tipo de enseñanza también en niveles educativos más elementales. Con el uso de esta tecnología, el docente puede crear aplicaciones para que el alumnado adquiera mejor las competencias en estos temas, además, estos pueden realizar sus propios experimentos y visualizar el cálculo de magnitudes como la varianza, la media o la función de distribución en función del número de eventos. Finalmente, el uso de nuevas tecnologías como javascript y html es discutido y evaluado.

**Palabras clave:** Monte Carlo, simulación, estadística, probabilidad, html, javascript, blog.

## 1. Introducción

Las matemáticas son consideradas, en las etapas educativas que comprenden la educación obligatoria, un área instrumental (junto a Lengua Castellana y Literatura, y Primera Lengua Extranjera), ya que el conocimiento que el alumnado adquirirá en dicha área o asignatura es esencial, para la formación pluridimensional del alumnado, y garantiza un correcto desarrollo en su futura vida personal, académica y laboral. Además, el "Conocimiento Matemático" está recogido en el listado de Competencias Básicas que recoge la actual Ley Orgánica 2/2006, de 3 de mayo, de Educación (LOE), entendiéndose las Competencias Básicas, como aquellos conocimientos y habilidades que el alumnado debe adquirir al finalizar la Educación Secundaria Obligatoria para integrarse de una manera adecuada en la sociedad. En concreto en Andalucía, el área de matemáticas, se incluye el siguiente núcleo temático (BOJA 171): *6. Tratamiento de la información, azar y probabilidad.*

Durante los primeros años del proceso educativo se deben aprender los conceptos de azar y probabilidad, junto a otros más complejos. La principal finalidad de este núcleo temático es que el alumnado comience a interpretar los fenómenos ambientales y sociales de su entorno cercano mediante en términos matemáticos. Esta debe entenderse como una disciplina que ayuda a interpretar la realidad y a actuar sobre ella de forma responsable, crítica y positiva. Para esa interpretación, juegan un papel esencial la estadística y a la probabilidad, disciplinas matemáticas entre las que existe una relación complementaria. En la actualidad, múltiples aplicaciones de dichas disciplinas invaden prácticamente todos los campos de la actividad humana y gozan de amplio reconocimiento social. Este hecho es constatado por su creciente importancia en el aprendizaje de otras materias, su presencia en los medios de comunicación general, en el mercado laboral y en el ambiente cultural. Por este motivo, el aprendizaje en todos los niveles educativos se inserta como una imprescindible meta de carácter cultural que ha de iniciarse de manera natural desde la educación primaria.

El desarrollo gradual comenzará, en los primeros cursos, por las técnicas para la recogida, organización y representación de los datos a través de las distintas opciones como tablas o diagramas, para continuar, en cursos sucesivos, con los procesos para la obtención de medidas de centralización y de dispersión que les permitan realizar un primer análisis de los datos. Al igual que para otros contenidos del área es recomendable la utilización del ordenador y de las calculadoras, tanto convencionales como gráficas, para manipular, analizar y representar conjuntos de datos. Los juegos de azar proporcionan ejemplos que permitirán introducir la noción de probabilidad y los conceptos asociados a la misma.

Por otro lado, la enseñanza de la estadística es altamente no trivial. Una de las principales dificultades es la formación específica de los profesores en este ámbito específico. En el artículo "¿Hacia dónde va la educación estadística?", Batanero (2000), afirma: "Los profesores que provienen de la Licenciatura de Matemáticas no tienen una formación especifica en didáctica de la estadística y muchos de ellos tampoco en estadística aplicada. La situación es aun peor en lo que se refiere a los profesores de primaria, la mayor parte de los cuales no han tenido una formación ni siquiera básica ya no sobre la didáctica de la estadística, sino sobre los conceptos básicos de estadística o probabilidad."

Otra dificultad es la escasez de herramientas que un profesor puede ofrecer a los alumnos para que estos desarrollen sus destrezas. Esta escasez provoca que a menudo la educación en estos temas se base en resolver problemas mediante la aplicación sistemática de un cierto procedimiento. En muy pocas ocasiones se puede proveer a los estudiantes de datos reales y de medios para poder analizarlos apropiadamente. Recientemente, la integración de las nuevas tecnologías, como hojas de cálculo, ha contribuido a socavar esta segunda dificultad, permitiendo realizar los cálculos de una manera más rápida y dejando más tiempo para los conceptos estadísticos.

Según la referencia Mills (2002), muchos investigadores en educación apoyan la teoría de que los estudiantes aprenden más eficientemente de forma activa, mediante la construcción de su propio conocimiento y el sentido que se le otorga al mismo. El constructivismo, Piaget (1896), sugiere que el nuevo conocimiento no se recibe pasivamente del maestro al estudiante a través de los libros de texto y conferencias, o simplemente pidiendo a los estudiantes memorizar las operaciones rutinarias. En su lugar, el significado se adquiere a través de una interacción significativa con los nuevos conocimientos, ver también von Glasersfeld (1987).

A pesar de la claridad con la que un profesor explica un concepto, los alumnos y alumnas comprenderán el material sólo después de haber construido su propio significado. Este proceso puede requerir la reestructuración y reorganización de nuevos conocimientos y su vinculación con el conocimiento previo o anterior. El constructivismo sugiere también que el aprendizaje debe ser facilitado por el profesorado y que la interacción y el debate son componentes críticos durante el proceso de aprendizaje, Eggen (2001).

La aplicación de los conceptos estadísticos, utilizando las simulaciones Monte Carlo, puede beneficiar a los estudiantes al animarlos a desarrollar su propia comprensión de los conceptos estadísticos. Los estudiantes tendrán la oportunidad de aprender mediante la construcción de sus propias ideas y el conocimiento de las experiencias de simulación por ordenador, con la dirección de apoyo por parte del instructor. Según Packard (1993) los estudiantes que participan activamente en su propio aprendizaje suele convertirse en aprendices más independientes y solucionadores de problemas.

Una ventaja interesante de las nuevas tecnologías, que se ha sugerido en la literatura, reside en su capacidad de mejorar la comprensión del alumnado de conceptos abstractos (Kersten (1983); Dambolena (1986); Gordon y Gordon (1989); Shibli (1990); Kalsbeek (1996); Hesterberg (1998)). Mediante el uso de la tecnología informática actual, es posible complementar el análisis de datos, con experiencias adicionales estadísticos mediante la utilización de métodos de simulación por ordenador. Estas simulaciones ofrecen a los estudiantes experimentar con muestras aleatorias de una población con parámetros conocidos, con el fin de clarificar conceptos abstractos y difíciles de la estadística.

En este artículo pretendemos mostrar cómo el docente puede usar herramientas habituales para la creación de páginas webs (HTML y JavaScript) para desarrollar las simulaciones necesarias para que el alumnado pueda realizar sus propios experimentos estadísticos. Mediante estas herramientas, el alumnado podrá participar en la construcción de su propio conocimiento con la dirección de apoyo por parte del equipo docente. Según Packard (1993), los estudiantes que participan activamente en su propio aprendizaje suele convertirse en aprendices más independientes y solucionadores de problemas. Mediante estos métodos, conceptos difusos como la media, mediana, varianza o percentil pueden ser calculados directamente, desarrollando así una mejor intuición al respecto.

2. Metodología.

Un blog, o bitácora, es una publicación web de historias que son presentadas en un orden cronológico. En ellos se pueden incluir texto, ilustraciones, enlaces a otras páginas web, o animaciones entre otras aplicaciones. También permiten la realización de comentarios por parte de los lectores, fomentando el debate entre el autor y la comunidad a la que va dirigido el blog.

En los últimos años, distintos editores de blogs se han popularizado, como Wordpress o Blogspot, permitiendo el uso de este medio sin la necesidad de unos elevados conocimientos informáticos. Por otro lado, es de gran utilidad familiarizarse con el lenguaje informático subyacente a esta nueva tecnología. El lenguaje más popular para la programación web se denomina HTML, siglas de HyperText Markup Language («lenguaje de marcado de hipertexto»), y se utiliza para describir y traducir la estructura y la información en forma de texto, así como para complementar el texto con objetos tales como imágenes. Otra útil herramienta para el diseño web es JavaScript. Este es un lenguaje que se añade al HTML y permite ejecutar instrucciones complejas en la web.

El uso de blogs en la docencia es un tema muy activo en la actualidad. En un reciente artículo, Lara (2005), se expresa que la versatilidad del formato blog como herramienta de gestión y publicación de contenidos ofrece múltiples posibilidades de uso educativo, y que cada vez más docentes van descubriendo y experimentando en sus respectivas áreas curriculares. Esta flexibilidad, junto con su sencillez de manejo gracias a las plataformas antes mencionadas, permite que se puedan adaptar a cualquier disciplina, nivel educativo o metodología. Un reciente ejemplo del buen uso de los blogs en la docencia de las matemáticas se encuentra en el blog Grima, C, "Mati y sus mateaventuras", ganador del premio bitácoras 2011 en la categoría de mejor blog educativo y en la de mejor blog.

Según Roschelle (2000) (ver también Hernández (2008)), mediante el uso de las nuevas tecnologías, y en concreto mediante los blogs, se pone de manifiesto la relación entre el aprendizaje significativo y el constructivismo. Mediante el uso de estas técnicas se demuestra que el aprendizaje es más significativo mientras estén presentes las siguientes características fundamentales: ambientes de aprendizaje actualizados y actualizables, compromiso activo, trabajo colaborativo, conocimiento del contexto real, y finalmente, la interacción frecuente y retroalimentación.

Javascript puede aportar a los blogs dinamismo e interacción. Un alumno puede aprender mejor si él mismo experimenta el concepto que se trate. Con Javascript el docente puede crear herramientas que permitan al alumno acercarse mejor a los conceptos. Las principal ventajas que ofrece este lenguaje es su facilidad de aprendizaje, junto con la cantidad de librerías que se pueden usar para construir nuevas herramientas. Con la aparición de HTML5 la representación gráfica ha evolucionado permitiendo el uso de gráficos en 3D.

Otra ventaja de esta aproximación es que es independiente del sistema operativo del usuario. Se verá correctamente si se usa el mismo navegador web sin importar si el sistema operativo es GNU/Linux, Microsoft Windows o MacOsX. Muchas de las herramientas que los alumnos encuentran en internet se basan en aplicaciones que están alojadas en un servidor y que se ejecutan allá donde éste se encuentre alojado. Con HTML y JavaScript, las herramientas se pueden usar incluso cuando no se tenga conexión a internet.

Por otra parte, las técnicas Monte Carlo son una herramienta muy usada en la investigación científica actual, Metrópolis (1987). La pieza clave de esta técnica es la producción de números pseudoaleatorios mediante algoritmos que pueden ser computados muy eficientemente con los ordenadores actuales. Mediante el algoritmo de Metrópolis, Metrópolis (1953), se pueden generar números aleatorios siguiendo una distribución de probabilidad arbitraria. El campo de aplicaciones de estas técnicas van desde la simulación de procesos físicos, la integración de funciones multivariable o la minimización. Dadas estas aplicaciones las técnicas Monte Carlo suelen formar parte del currículum de carreras científicas, como física o química, véase por ejemplo la web del profesor Pedro Garrido de la universidad de Granada. La aplicación, por otro lado, en educación primaria o secundaria es mucho menor.

3. **Resultados.**

En el artículo Weir (2002), estudian el comportamiento en cuanto el aprendizaje de varios grupos de alumnos frente a aplicaciones desarrolladas para mostrar distintas distribuciones de probabilidad. Se nos muestra que el uso de estas aplicaciones basadas en técnicas de simulación Monte Carlo favorece el aprendizaje de los estudiantes. Para ser más precisos, encuentran un aumento significativo de aprendizaje en el alumnado que presentaban dificultades con otros métodos más clásicos.

En nuestro trabajo hemos desarrollado varias aplicaciones que pueden ser útil para facilitar el aprendizaje en estadística. Una de ellas explora la falacia del jugador, un resultado aparentemente contra intuitivo que suele ser difícil de asimilar por los alumnos. Esta supuesta contradicción aparece cuando una persona espera un resultado concreto para un siguiente dato vista la historia de los anteriores, para variables sin correlación temporal. Para ilustrar este fenómeno simulamos el lanzamiento de N monedas, mediante la generación de números pseudoaleatorios, y repetimos este lanzamiento varias veces hasta que salgan todas con el mismo estado, todas caras o todas cruz. Una vez realizado esto se entrega la información al usuario y preguntamos qué va a salir en el siguiente lanzamiento. Tras realizar la experiencia repetidas veces el usuario puede asimilar que el siguiente resultado de la moneda no es dependiente de los anteriores. La aplicación interactiva se puede encontrar en el blog de Daniel Manzano.

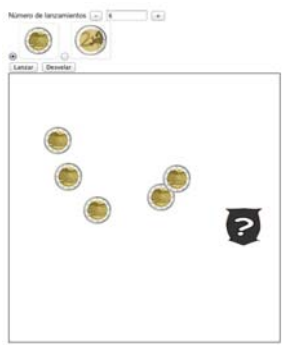

Figura 1: Aplicación de la falacia del jugador.

En otra aplicación se puede observar el Teorema del Límite Central. Según este teorema clásico, la suma de N números aleatorios independientes, con una media η y varianza σ finitas, se comporta como un elemento de una distribución gaussiana con media Nη y varianza $\sigma/\sqrt{N}$.

Usando el algoritmo de Metrópolis se pueden generar números aleatorios que sigan unas distribuciones de probabilidad. Nuestra aplicación genera un número N de variables siguiendo una determinada distribución y calcula su suma. La aplicación una vez obtenida una muestra de datos de esa suma dibuja un histograma reflejando que a medida que se aumenta el número de eventos se acerca más a una distribución gaussiana. Así el alumnado puede relacionarse con el significado real del Teorema del Límite Central, que es una pieza clave en todas las ciencias empíricas. También mediante la generación progresiva del histograma se puede comprender la diferencia entre tener mucha o poca estadística, y el alumnado puede entender la diferencia entre tener muchos o pocos eventos. Finalmente, mediante esta aplicación se pueden simular distintos experimentos relacionados con la física, la química o la economía, transmitiendo conceptos importantes como las barras de error o los percentiles. Estos "experimentos virtuales" pueden ser comparados por el docente con experimentos reales y actuales como los realizados en el CERN, muchos de cuyos datos son públicos, véase la web del CERN.

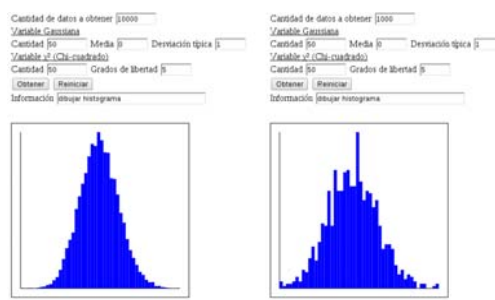

Figura 2: Histogramas generados con la aplicación del Teorema del límite central, con 10000 datos (izquierda) y 1000 (derecha)

Finalmente, hemos desarrollado otra aplicación donde se muestra la utilidad que puede tener la estadística y la probabilidad en campos a priori ajenos a la materia, como la geometría. El objetivo de la aplicación es calcular el valor de la constante π. El método usado es generar números aleatorios dentro de un cuadrado que a su vez contiene un círculo. Mediante el cálculo de la proporción de números dentro y fuera del círculo el área relativa del círculo con respeto al cuadrado puede ser calculado, y de ahí el valor de la constante. La aplicación realiza el cálculo en tiempo real, a medida que va generando los puntos. De esta manera se puede apreciar como se acerca al valor real progresivamente. Esta aplicación se encuentra alojada en el blog de F.M. Alexander Bueno.

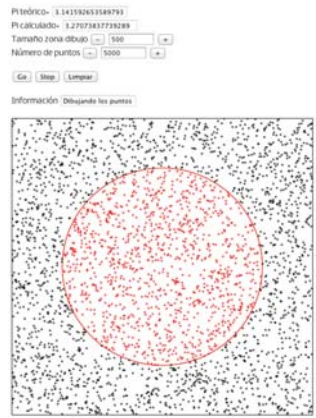

Figura 3: Aplicación del cálculo del área de un círculo.

Todas estas herramientas están licenciadas como software libre, lo cual permite a los profesores modificar y usar según sus necesidades. Esta opción, buscar licencias abiertas, no es arbitraria. Mediante el uso de licencias de software libre se permite al resto de usuarios modificar, documentar o adaptar libremente las aplicaciones disponibles, facilitando además la creación de una comunidad docente que colabore en esta dirección.

## 4. Conclusiones

En este artículo hemos presentado una técnica matemática, las técnicas Monte Carlo, y técnicas informáticas, HTML y Javascript, como herramientas útiles para la enseñanza de la estadística y la probabilidad en la enseñanza primaria y secundaria. Mediante el uso de estas nuevas técnicas es posible realizar experimentos virtuales que acercan al alumnado conceptos abstractos y permiten desarrollar la intuición al respecto. Estos experimentos virtuales pueden ser transmitidos mediante blogs, de modo que el alumnado pueda acceder al contenido fuera del horario docente y practicar. Finalmente, si las simulaciones se licencian como software libre, el alumnado puede participar de la modificación y creación de nuevas herramientas, fomentando así el aprendizaje constructivo.